# Perfect Secrecy Using Compressed Sensing

Mahmoud Ramezani Mayiami, *Student Member, IEEE*, Babak Seyfe, *Senior Member, IEEE*
Hamid G. Bafghi, *Student Member, IEEE*,

*Abstract*—In this paper we consider the compressed sensing-based encryption and proposed the conditions in which the perfect secrecy is obtained.

We prove when the Restricted Isometery Property (RIP) is hold and the number of measurements is more than two times of sparsity level i.e. $M \geq 2k$, the perfect secrecy condition introduced by Shannon is achievable if message block is not equal to zero or we have infinite block length.

## I. Introduction

In the recent years, many researchers have used the theory of Compressed Sensing or Compressive Sampling (CS) for their field of study. Candes and his colleagues in [1] and Donoho in[2] proposed this theory and it has been served in some applications such as image processing[3],radar [4], signal detection [5]and other applications [6].

Achieving the perfect secrecy to have secret communication is a challenging problem in the field of information theory. Information-theoretic secrecy intends to ensure that an eavesdropper who listens to the wireless transmission of a message can only collect an arbitrarily small number of information bits about this message and this fact was introduced by Shannon in his fundamental paper [7]. He tried to minimize the leakage of information to those unintended receivers, i.e. eavesdroppers, and explained the *perfect secrecy condition* in which the listening to the channel cannot increase the probability of decryption of the sent message for the eavesdropper.

In the classical secret communication's approach, the messages encrypted and compressed, separately. But, recently, encryption was complied by compressed sensing, e.g. [8] and [9]. In this approach, a measurement matrix which is generated for compressive sampling purpose is used as a key to encrypt the sparse messages at the same time.

Rachlin and Baron in [8] argued whether the measurements of compressed sensing can prepare an approach to obtain perfect secrecy simultaneously or not. They investigate the achievability of perfect secrecy by using the measurement matrix of compressive sampling as a key for encrypting the transmitted signal and they proved compressed sensing-based encryption cannot achieve perfect secrecy.

In this paper we consider the perfect secrecy problem in compressed sensing measurements and introduce some special conditions in which the perfect secrecy via compressed sensing is achievable. Although it is called as special cases, they have many practical usages.

The organization of this paper is given in the following. In Section II, we review the literature of compressed sensing and perfect security. Perfect secrecy through compressed sensing, as the main idea of this paper, is proposed in Section III. In SectionIV, we will argue about some resemblances and conflicts between our results and the results presented at [8]. We will conclude this paper in section V.

## II. The Backgrounds

In what follows we introduce some preliminaries about compressed sensing and security.

### A. Compressed Sensing

By CS approach, we can sense the signal in compressed form. In other words, the important information of signal will be sensed and the reminder that could not be useful will be ignored.

Suppose $x$ is a $N \times 1$ signal vector that we want to find its elements. It is shown that $x$ can be interpreted as $x = \Psi \alpha$, where $\Psi$ is a specific $N \times N$ dictionary that its columns are orthonormal and spans $x$ domain and $\alpha$ is the coefficient vector of $x$ in basis $\Psi$. The vector $\alpha$ is said to be $k$-sparse whenever $k$ non-zero elements exist in $\alpha$ for $k \ll N$. Here, we count non-zero elements with $l_0$-norm notation i.e., $\|\alpha\|_0 = k$.

In the compressed sensing, instead of sensing the signal directly, we observe $M$ measurments of signal vector ($k < M < N$) and these measurement scan be shown by $y$ as

$$y = \Phi x = \Phi \Psi \alpha = A \alpha. \quad (1)$$

Here, $\Phi$ is $M \times N$ measurement matrix and $A$ is holographic dictionary [2].

Since $M < N$, (1) is underdetermined system of linear equations which has many solutions. If we choose a constraint on sparsity level of $\alpha$, we can be assure of unique solution as following

$$\min_{\alpha} \|\alpha\|_0 \text{ subject to } y = A\alpha. \quad (2)$$

There are some approaches for implementing (2) such as Orthogonal Matching Pursuit or OMP algorithm [10]-[11]. In addition, if $A$ has Restricted Isometry Property (which will be defined latter), we can recover $\alpha$ as[12]

$$\min_{\alpha} \|\alpha\|_1 \text{ subject to } y = A\alpha, \quad (3)$$

The authors are with the Department of Electrical Engineering, Shahed University, Tehran, Iran, Email: seyfe@shahed.ac.ir.
This work was supported in part by IRAN Telecommunication Research Center (ITRC), under Grant numbers T/500/4800 and T/500/3133.

where $\|\alpha\|_p$ indicates $l_p$-norm which is defined as $\|\alpha\|_p = \sqrt[p]{\sum_{i=1}^{N}|\alpha_i|^p}$ and $\alpha_i$ for $1 < i < N$ are the elements of vector $\alpha$. Basis Pursuit (BP)[13] is one of the approaches for computing (3).

*Definition 1* [12]: $A$ respects Restricted Isometry Property (RIP) of order $k$ when for all $k$-sparse vector $\alpha$ with appropriately chosen constant $0 < \varepsilon_k < 1$, $\|A\alpha\|_2$ satisfies constraints as following

$$(1 - \varepsilon_k)\|\alpha\|_2 \leq \|A\alpha\|_2 \leq (1 + \varepsilon_k)\|\alpha\|_2. \quad (4)$$

However an important question is: what measurement matrices (with respect to specific dictionary $\Psi$) does make $A$ to satisfy RIP?

The authors in [14] proved that if the elements of $\Phi$ are selected from independent and identically distributed (i.i.d) random variables from a Gaussian probability density function (pdf) with mean zero and variance $1/M$, then $\Phi$ will be incoherent with any basis $\Psi$. So $A$ satisfies RIP with overwhelming probability for $M \geq ck\log(N/k)$, with some constant $c$ [1], [2], [15].

*B. Perfect Secrecy*

Perfect secrecy which is introduced by Shannon in his fundamental paper [7]. It is based on the statistical properties of a system, and provides protection even in the face of a computationally unbounded adversary. In the Shannon's model, a source message $W$ is encrypted to a ciphertext $E$ by a key $K$ that is shared by the transmitter and the receiver. An eavesdropper, which knows the family of encryption functions (keys) and the probability of choosing keys, may intercept the ciphertext $E$. The system is considered to be *perfectly secure* if *a posteriori* probabilities of $W$ for all $E$ would be equal to *a priori* probabilities independent of the values of $E$, i.e., $P_{W|E} = P_W$. Alternatively, this condition can be stated as $I(X;Y) = 0$, where $X$ is the transmitted message and $Y$ is the received signal in eavesdropper. In addition, Shannon proved the pessimistic result that perfect secrecy can be achieved only when the secret key is at least as long as the plaintext message or, more precisely, when $H(K) \geq H(W)$.

In the sequel, we try to achieve perfect secrecy through the compressed sensing measurements.

III. THE MAIN IDEA

Suppose $X$ and $Y$ are discrete random variables with alphabet $\mathcal{X}$ and $\mathcal{Y}$, respectively and $\mathcal{X}$ and $\mathcal{Y}$ contains source messages and encrypted messages, respectively. Each of the source messages $x \in \mathcal{X}$ is sparse in the basis $\Psi$ and for the sake of simplicity and without loss of generality, suppose that $\Psi$ is an identity matrix. Also we assume that the channel is not noisy and the transferred signal is not interfered. Compressed sensing-based encryption expresses that we can transmit cryptogram $y = \Phi x$ instead of $x$ and in the receiver, we can decrypt it with knowledge of $\Phi$. Hence, the eavesdropper receives the encrypted message exactly i.e. $y = \Phi x$.

The measurement matrix $\Phi$ can be selected from a set of keys that is known for the transmitter and the permitted receiver. Each random measurement matrix $\Phi$ is generated with a seed which can be exchanged through a secure approach between two desired sides [16], [17].

In the following theorems, we will show that the compressed sensing based-encryption satisfies Shannon's definition of perfect secrecy when some conditions were satisfied

*Theorem 1:* Let $X$ has a uniform distribution over $\mathcal{X}$. The compressed sensing based-encryption achieves perfect secrecy if:

i. the number of measurements M is equal or greater than two times of sparsity level of the messages, i.e., $M \geq 2k$,
ii. the measurement matrix $\Phi$ satisfies RIP and
iii. the number of source messages goes to infinity.

*Proof:* To compute the mutual information $y = \Phi x$ we have

$$\begin{aligned}
I(X,Y) &= H(Y) - H(Y|X) \\
&= H(Y) \\
&\quad - \sum_{x \in \mathcal{X}} H(Y|X = x)P_X(X = x) \\
&= H(Y) \\
&\quad - \Big\{ H(Y|X = 0)P_X(X = 0) \\
&\quad + \sum_{x \in \mathcal{X}, x \neq 0} H(Y|X = x)P_X(X = x) \Big\} =^{(a)} H(Y) \\
&\quad - \sum_{x \in \mathcal{X}, x \neq 0} H(Y|X = x)P_X(X = x) =^{(b)} \log|T| \\
&\quad - \sum_{x \in \mathcal{X}, x \neq 0} H(Y|X = x)P_X(X = x) \\
&=^{(c)} \log|T| \\
&\quad - \frac{1}{T}\sum_{x \in \mathcal{X}, x \neq 0} H(Y|X = x) \\
&= \log|T| - \frac{T-1}{T}\log|T-1|.
\end{aligned} \quad (5)$$

where

$(a)$ is because of the fact that for message $x$ equal to zero, then $y = 0$ and $H(Y|X = 0) = 0$.

$(b)$ regarding the facts that if $M \geq 2k$, every message $x$ has a unique projection [18] and uniform distribution is supposed over the source messages in the theorem, $Y$ has a uniform distribution over $T$ cryptograms in $\mathcal{Y}$, hence $H(Y) = \log|T|$.

$(c)$ are because that $Y$ has a uniform distribution over $\mathcal{Y}$, therefore we have $P_X(X = x) = \frac{1}{T}$.

It is clear that when the number of source goes to infinity (i.e., $T \rightarrow \infty$), then $I(X,Y) = 0$ and the perfect secrecy will be achieved.

*Theorem 2:* Suppose that $M \geq 2k$, $\Phi$ satisfies RIP and $X$ has a uniform distribution over the $\mathcal{X}$. If there isn't any null message in the set of source messages, i.e., $\forall x \in \mathcal{X}, x \neq 0$, perfect secrecy will be achieved via compressed sensing.

*Proof:* Since $\Phi$ satisfies RIP, there isn't any source messages in the null space of $\Phi$ [19] and because it is assumed that the null message doesn't exist, i.e., $x \neq 0$, hence we do not have cryptogram $y = 0$. Also, since $M \geq 2k$, every message $x$ has a unique projection [18] and then, each messages of $\mathcal{X}$ will be encrypted to unique cryptogram that belongs to $\mathcal{Y}$. Hence $Y$ has a uniform distribution over $\mathcal{Y}$ and $H(Y|X) = \log|T|$ and perfect secrecy achieves as

$$I(X,Y) = H(Y) - H(Y|X) = \log|T| - \log|T| = 0. \quad (6)$$

## IV. COMPARING WITH THE PREVIOUS WORK

Here, there are some resemblances and conflicts between our results and results of [8]. It was reported in [8] that compressed sensing-based encryption have notability to obtain Shannon's definition of perfect secrecy in general. To prove this postulate, they use two proofs as following.

The first or original proof expresses that $P_{Y|X}(Y = 0|X = 0) \neq P_Y(Y = 0)$ and then $X$ and $Y$ are dependent, hence $I(X,Y) > 0$, which means that perfect secrecy is not achievable.

Note that in general cases the perfect secrecy is not achievable. Since in our first theorem $T \to \infty$, $I(X,Y)$ approximates to zero and is not zero exactly. Also in the second theorem we have shown when the null message is eliminated from the alphabet $\mathcal{X}$, the perfect secrecy can be achievable and this postulate doesn't conflict with the first proof in [8].

In the second proof in [8], Rachlin and Baron demonstrate that if a message such as $x' \epsilon \mathcal{X}$ is chosen such that $\| x' \|_2^2 < \| y \|_2^2 / (1 + \varepsilon_K)$, then we will have $P_{Y|X}(X = x'|Y = y) = 0$ and they assumed nonzero a priori probability for this message, i.e., $P_X(x') > 0$. Hence because of $P_{Y|X}(X = x'|Y = y) \neq P_X(x')$, the perfect secrecy through compressed sensing is not possible.

We would like to draw attentions to that if $\| x' \|_2^2 < \| \Phi x \|_2^2 / (1 + \varepsilon_K)$ then $\Phi$ does not satisfy RIP. In other words, if we assume that $\Phi$ satisfy RIP, then the above assumption is not valid for $x'$, see (4), and this message does not exist in the set of source messages $\mathcal{X}$. Hence $P_X(x') = 0$ and $P_{Y|X}(X = x'|Y = y) = P_X(x')$ and we can not contend that perfect secrecy is not achievable.

On the other hand, they proved their lemma by contradiction in two special cases in which perfect secrecy is not achievable for compressed sensing measurement. They derived that this measurement cannot satisfy perfect secrecy condition. So we presented a new theorem in which one of their counterexample is not included in message set.

## V. CONCLUSION

In this paper, the perfect secrecy via compressed sensing was studied. It is shown that with the uniform distribution over source messages and specific restriction on the number of measurements, compressed sensing-based encryption can achieves the Shannon's definition of perfect secrecy.